\documentclass[letterpaper,twocolumn,10pt]{article}
\usepackage{usenix-2020-09}

\usepackage{tikz}
\usepackage{amsmath}

\begin{document}

\date{}

\title{\Large \bf Exploring and Improving the Accessibility of Data Privacy-related Information for People Who Are Blind or Low-vision}

\author{
{\rm Yuanyuan Feng\textsuperscript{1}, Abhilasha Ravichander\textsuperscript{2},Yaxing Yao\textsuperscript{3}, Shikun Zhang\textsuperscript{2}, Norman Sadeh\textsuperscript{2}}\\
University of Vermont\textsuperscript{1}, Carnegie Mellon University\textsuperscript{2}, University of Maryland Baltimore County\textsuperscript{3}
}



\maketitle

\begin{abstract}
We present a study of privacy attitudes and behaviors of people who are blind or low vision. Our study involved in-depth interviews with 21 US participants. The study explores their risk perceptions and also whether and how they go about obtaining information about the data practices of digital technologies with which they interact. One objective of the study is to better understand this user group's needs for more accessible privacy tools. We also share some reflections on the challenge of recruiting an inclusive sample of participants from an already underrepresented user group in computing and how we were able to overcome this challenge.
\end{abstract}


\section{Introduction}
Accessing information about the data practices of technologies such as the text of privacy policies or available privacy choices is not always easy. For one, finding privacy policies, the legal documents mandated by many privacy laws around the world, is often not as straightforward as one might expect~\cite{sundareswara2020privacy}.
Even if people manage to locate privacy policies, understanding the data practices and privacy implications disclosed in these documents can be even more challenging, as the texts of privacy policies are know to be long, vague, and filled with technical and legal jargon\cite{mcdonald2008cost, reidenberg2015disagreeable,krumay2020readability}.
In this work, we try to shed light on how people who are blind or low-vision navigate data-privacy related information, the specific challenges they encounter, and what can be done to assist them -- either through technology or through new standards or regulations.

According to Cornell University Employment and Disability Institute's interpretation of the 2016 American Community Survey, more than 7 million (2.4\% of the total US population) reported having a visual disability~\cite{BlindnessStats}.
Besides people who are blind, low vision caused by various levels of vision loss is also common among older adults~\cite{pelletier2016vision}.
In this paper, we use ``blind'' and ``low-vision'' (instead of ``visually-impaired'') to refer to this group of digital technology users, as they are positive affirming terms recommended the National Federation of the Blind (NFB)~.\footnote{The National Federation of the Blind (\url{https://nfb.org/})}

People who are blind or low-vision consist of a nontrivial user group of various digital technologies, who often rely on assistive technologies (e.g., screen readers) to access other digital technologies and services. 
Research shows that they are at higher risks of online privacy and security threats due to the lack of visual cues~\cite{ahmed2015privacy, hayes2019cooperative}, while generic technologies tend to have poor accessibility features to support their information needs in the digital world~\cite{dosono2015m, kim2016interaction}.
Therefore, greater research efforts are necessary to improve not only the usability but also the accessibility and inclusiveness for this user group.

In this study, we set out to understand this user group's challenges and needs in navigating data privacy-related information of various digital technologies. To develop a comprehensive understanding in this topic space, we also need to explore user group's attitudes and perceived risks around data privacy to establish a baseline, which may differ from those of sighted users.
In addition, this study serves as the formative research for potential technical solutions that could assist this user group in navigating data privacy-related information. 
In this position paper, we first articulate the motivation of our recent fully remote in-depth interview study with 21 US participants who are blind or low-vision. Then, we describe our study methods and our remote data collection process during 2021 in the pandemic context. Finally, we share our reflections on inclusive recruitment strategies from this study.

\section{Related Work}
With the growing recognition of the importance of accessibility and inclusiveness in privacy and security technologies~\cite{wang2018inclusive}, many research studies tried to understand blind and low-version users' privacy concerns and behaviors.\footnote{When describing prior research, we consistently use our preferred terms (i.e., blind, low-vision) despite the terms used in the original publications.} 
Ahmed et al.~\cite{Ahmed2015}'s interview study revealed blind participants' privacy concerns in three environments: physical (e.g., eavesdropping), digital (e.g., privacy settings in social media), and the intersection of physical and digital (e.g., shoulder-surfing). This work highlighted privacy risks at the intersection of physical and digital space that are unique to this population. Hayes et al. found that this population relied heavily on their ``allies'' (e.g., family members, caregivers) to cooperatively protect their privacy and security~\cite{hayes2019cooperative}. 
Moreover, Akter et al.’s survey study examined this population's concerns regarding camera-based assistive technology and services, where volunteers answer their questions about photos or videos~\cite{akter2020uncomfortable}. 

Another important body of work focuses on the usability and accessibility of security and privacy tools for people who are blind or low-vision. For example, Dosono et al. found that web authentication can be very time-consuming for this user group and pose significant challenges, such as accessing error messages~\cite{dosono2015Im,Dosono2018}. More importantly, such usability issues may lead members of this group to resort to risky behaviors or make decisions that compromise security~\cite{napoli2021m}. Another study of this population's online information behaviors revealed the barriers faced by blind web users when it comes to assessing the credibility of websites~\cite{abdolrahmani2016should}.

Concurrently, computer scientists have also started to explore new technical solutions and novel interaction modalities to improve the accessibility of digital information for people who are blind or low-vision, such as computer vision~\cite{Gurari_2018_CVPR} and voice-based personal assistants~\cite{Pradhan10.1145/3173574.3174033}.
Considering the predominantly textual nature of data privacy-related information online, technologies leveraging natural language processing (NLP) could prove promising to assist people who are blind or low-vision navigating such information.
Our group's prior research includes using NLP techniques to analyze the text of privacy policies~\cite{wilson2018analyzing}, identify and categorize different data practice disclosures~\cite{liu2018towards,sadeh2013usable}, and also use this analysis to automatically answer people's privacy questions~\cite{ravichander2019question, ravichander2021breaking}. Therefore, we are also motivated to explore the feasibility of a digital assistant that can answer users' data privacy questions based on publicly available privacy policies, as an accessibility tool to assist people who are blind or low-vision in navigating data privacy-related information.

\section{Pseudo-hypotheses and Research Questions}
Based on our literature review and consultation with two professional contacts who are blind, we developed two pseudo-hypotheses for this study: First, people who are blind or low-vision are likely to face challenges that are different from sighted users, in finding data privacy-related information. Second, they are in need of more accessible privacy tools to assist them in finding and/or understanding such privacy-related information.

We were aware that these pseudo-hypotheses may or may not be true, which motivated us to choose the in-depth qualitative interviewing method to explore this research topic without making assumptions. This allowed us to approach the topic in a relatively conversational and open-ended manner. Our consultants also gave us valuable advice to finalize our research questions without making insensitive assumptions. For example, we we advised to be mindful of this group's varying levels of familiarity with different digital technologies, as well as the high unemployment rate of this population in the US~\cite{Unemployment}.

To explore our two pseudo-hypotheses, it is critical to establish a baseline understanding of this user group's attitudes and perceived risks associated with privacy regarding the digital technologies with which they interact. Accordingly, our study aims to answer the following three research questions (RQs).

RQ1: What are the attitudes and perceived risks associated with data privacy among people who are blind or low-vision?

RQ2: What are the information behaviors of data privacy-related information of people who are blind or low-vision?

RQ3: What are the needs (if any) of people who are blind or low-vision for assistive technologies to navigate data privacy-related information?

\section{Study Protocol and Data Collection}
In this section, we describe our study protocol and data collection for this study, which was conducted in 2021. We also reflect on the lessons learned by working with this group of participants. This study protocol was approved by the Institution Review Board (IRB) at Carnegie Mellon University. The research data is also governed by a data use agreement after the first author moved to her current institution.

\subsection{Study Protocol and Interview Questions}
We finalized our study protocol after seeking advice from two consultants and a pilot interview with a personal contact, all of whom are blind. 
Before the start of the study, we asked participants about their preferred vocabulary to describe their vision status (or disability) so that we can use their preferred vocabulary consistently throughout the interview. Then, we moved on to obtain their verbal consent and then started the interview and audio recording. 
We structured our interview questions around the three research questions, with additional technology use questions at the beginning and optional demographic questions at the end. We describe the rationale for developing the study protocol and interview questions in this section. The Appendix B shows the key interview questions, and we plan to make all questions available in our upcoming full paper.

\subsubsection{Technology use}
Following our consultants' advice, we situated our research questions in participants' diverse digital technology use. Therefore, the interview protocol starts with a set of general questions asking participants to describe their daily digital technology use practices.
After they reported their technology use, we  provided them with a plain language explanation of the terms such as ``digital technologies'' used in the interview questions to ensure terminology consistency (Appendix A).

\subsubsection{Attitudes and perceived risks} 
The second set of questions aimed to explore participants' attitudes and perceptions around data privacy, primarily on the digital technologies that they reported using earlier.
After participants explained their understanding of data practices pertaining to the digital technologies they use, we provided plain language explanations for ``data privacy'' and ``data practices'' in relation to their reported understanding to ensure the consistency of terminology (Appendix A).
Furthermore, the last question in this set used the critical incident techniques to bring out participants' memorable experiences~\cite{flanagan1954critical}.
We purposefully did not exclude other privacy issues that participants may want to share, since the intersection of digital and physical environments is known to be another source of data privacy concerns for this user group~\cite{Ahmed2015}. 

\subsubsection{Data privacy-related information behavior}
The third set of questions examined participants' information behavior with data privacy-related information of the digital technologies with which they interact.
We asked participants about how they navigate, access, and understand such information, while not assuming that they intentionally seek it.
For participants who reported having sought data-privacy related information before, we further asked about how did they go about finding such information and whether they experienced any difficulties or not.
We also explored their perceived credibility of different information sources since trust is a key factor in the communication of data privacy practices~\cite{waldman2018privacy}.

Our larger research project also aims to explore technical solutions to make privacy policies more usable for all. 
Though privacy policies are the standard mechanism for disclosing data practices, we did not assume that participants had any prior experience looking for privacy policies, let alone reading them. 
We started with general information behavior questions to see if they naturally bring up privacy policies as an information source. 
We waited until the last few questions to prompt them about their possible awareness and use of privacy policies if they did not mention privacy policies earlier in the interview protocol.

\subsubsection{Needs for assistive technologies to navigate data privacy-related information}
The final set of questions encourages participants to articulate their needs (if any) for assistive technologies to navigate information about data practices of different digital technologies with which they interact.
We purposefully did not assume our group's research direction (i.e., automatic answering data privacy questions based on privacy policies) is the ideal solution.
Instead, we started with a hypothetical scenario, asking participants what questions they would ask a data privacy expert about specific digital technologies. We selected up to four digital tools/technologies by category (i.e., website, mobile app, voice assistant, assistive technology) that each participant reported using earlier. 
Then, we ask about their opinions about the functionality our proposed solution could provide while not mentioning our solution. We wait until the last few questions to discuss their perceived pros and cons of a privacy tool in line with our research direction.

\subsubsection{Demographic questions (optional)}
At the end of the interview, we asked participants a standard set of demographic questions. We made it clear that they are optional, but it will be valuable data to ensure our study is inclusive. All participants chose to answer these questions.

\subsection{Recruitment and Data Collection}
Recruiting people who are blind or low-vision into research studies can be challenging and requires researchers to build trust among potential participants.
In our IRB protocol, we planned to recruit US participants from several channels, including our own contacts, local organizations, and the National Federation of the Blind, with the snowball sampling as a backup~\cite{handcock2011comment}. 

Our consultants advised us to be mindful of the sampling bias towards those who are already comfortable using assistive and other digital technologies. Therefore, we include the first author's phone number as a contact channel in non-public recruitment materials for those who prefer phone calls or text messages. Also, we used Zoom\footnote{Zoom (\url{https://zoom.us/})} to conduct the remote interviews, which has a one-tap dial-in function that may be preferred by some participants. Finally, we provided customized instructions on how to join a Zoom call in case participants had never used Zoom before.

To ensure diversity and inclusiveness of our sample, we explicitly insert a statement in all of our study recruitment materials: ``We particularly welcome diverse perspectives from individuals who are less familiar with technology and who also belong to other underrepresented groups.'' To achieve inclusiveness in recruitment, we did an informal screening by asking if they belong to any underrepresented groups when we received messages from potential participants. We politely expressed our rationale to include a diversity of participants in this study, which seemed to be well-received by many participants who were not chosen to participate. 

This study recruited 21 participants in the US who self-reported to be blind or low-vision, two of which were from snowball sampling and the remaining from the approved email solicitation distributed by the National Federation of the Blind. 
All remote interviews were audio recorded using Zoom's built-in functionality. Since some participants joined the interview via computers with videos on, we asked them to turn off their cameras during the interview to ensure that only audio was recorded.
The interviews ranged from 40 to 92 minutes in length (mean = 65 minutes) and each participant was compensated with an accessible electronic gift card of 25 US dollars.

We leveraged Zoom's automatic transcription functionality to generate the initial transcripts for all interviews. Then the research team went through all audio recordings again to correct the mistakes in the transcripts and re-structure them based on questions.
These clean transcripts are the primary qualitative data collected from this study, and we are currently in the process of analyzing the qualitative data.

\section{Reflections on Inclusive Recruitment}
In this section, we describe the diverse participants recruited into this study and reflect on inclusive recruitment strategies.

\subsection{A Diverse Sample}
Among 21 participants in this study, 19 self-reported being blind or legally blind and two of them reported being low-vision. Eleven and ten participants self-identified as female and male, respectively.

\begin{figure}
  \centering
  \includegraphics[width=8cm]{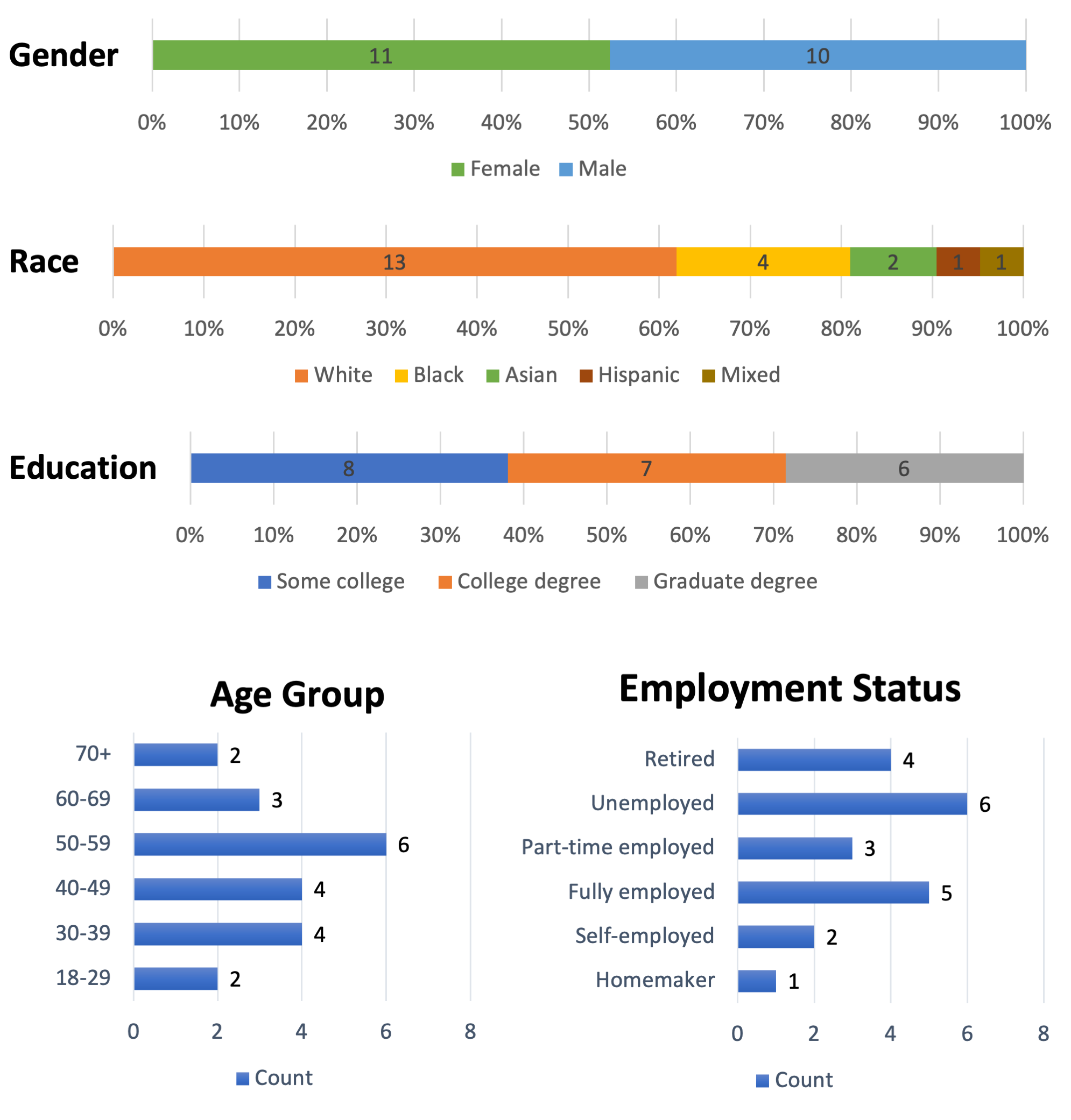}
  \caption{Basic Demographics of the Participants}
  \label{fig:demographics}
\end{figure}

Figure \ref{fig:demographics} shows basic demographics of our sample. We were able to over sample non-white participants (38\%) than the US demographic distribution.
Our participants also spread across all age groups and different employment statuses. 
Though we did not explicitly ask in our demographic questions, two female participants voluntarily shared with us that they belong to the LGBTQ+ community.
Our sample is biased towards people with relatively high education level, as all participants reported having at least some college education.
Nevertheless, for a qualitative research study with a relatively small sample size, we successfully recruited an inclusive sample among people who are blind or low-vision.

\subsection{Towards Inclusive Recruitment}
From our successful recruitment of an inclusive sample within an already underrepresented user group in computing within the challenging pandemic context, we want to share our thoughts on inclusive recruitment strategies.

First, recruiting people who are blind or low-vision into research studies already requires plenty of accessibility considerations.
At the early stage of the study, we were lucky to have the opportunity to obtain advice from two consultants who are also blind.
Their valuable advice helped us avoid common pitfalls and logistical issues that may not have occurred to us. For example, for a remote interview study, a waiver of written consent from IRB is a necessary accommodation for participants who may not have easy access to digital signing software.
Generally, for user studies that recruit from underrepresented groups, we recommend consulting someone who belongs to the group(s) early in the process.

Second, it is worth noting that we were unsuccessful with our recruiting efforts through local organizations. Probably as a result of the pandemic, the local organizations for people who are blind or low-vision either declined or did not respond to our inquiries.
We recruited two participants through snowball sampling from the pilot study participant. The majority of participants were recruited through the National Federation of the Blind, which has a strict reviewing process for research study solicitation targeting their members. 
Going through their reviewing process was helpful and also educated us about their preferred affirmative terms (e.g., low-vision).

Finally, ensuring inclusiveness when recruiting from an already underrepresented population is even more challenging, but we believe it is worth the effort.
We found it helpful to state the inclusive recruitment goal of our research study upfront: in our recruitment materials, in the study inclusion criteria, and before we ask potentially sensitive demographic questions. 
It is our observation that our messages were well received by a number of study participants and some potential participants who were not selected.

\section{Conclusion}
We presented the motivation, study design, and data collection of an in-depth interview study with 21 US participants who are blind or low-vision, exploring how to improve the accessibility for this user group to navigate data privacy-related information in the digital world.
Particularly, we reflected on our successful recruitment of an inclusive sample, which we hope to be valuable to other inclusive privacy and security researchers. 
We are in the process of analyzing the study data and intend to share our preliminary results from this study at the 7th Workshop on Inclusive Privacy and Security.



\section*{Acknowledgments}
We want to thank Chancey Fleet and Dr. Cynthia Bennett for their valuable advice on the study, as well as the National Federation of the Blind for helping us recruit an inclusive sample.
This research was supported in part by the National Science Foundation Secure and Trustworthy Computing program (19-14486) under a grant on Automatically Answering People's Privacy Questions and was also conducted under the broader umbrella of the ``Usable Privacy Policy Project'' (\url{https://www.usableprivacy.org}). Additional support for this research was also provided by a grant from DARPA and AFRL to the Brandeis project on Personalized Privacy Assistants (FA8750-15-2-0277; \url{https://www.privacyassistant.org})

\bibliographystyle{plain}
\bibliography{WIPS2022.bbl}

\appendix

\section{Explanation of terms to participants}
\noindent \textbf{Digital technologies}: During this interview, I will use the general term ``digital technologies'' to refer to all the websites, apps, web services, and assistive tools/technologies you mentioned. Does it sound okay to you?

\noindent \textbf{Data practices}: As you may know, when you use these digital tools/technologies on your devices, your personal data is often collected, used, or shared by these tools/technologies.

\noindent \textbf{Data privacy}: [You mentioned some privacy concerns with how your personal data is being handled by different digital tools/technologies] or [There are privacy concerns around how people’s personal data is being handled by different digital tools/technologies] or [Data privacy concerns the handling of personal data by different entities, such as if the handling is appropriate and if it’s in compliance with laws. The handling of personal data includes a variety of practices, such as how data is collected, used, or stored, whether data collectors share or sell the data to others.] This is the main research topic of this interview study, which is data privacy with digital technologies, or digital data privacy.

\section{Key Interview Questions}
\textbf{Technology use}
\begin{itemize}
\item What electronic devices do you typically use to access digital information?
\item What assistive tools or technologies and other websites/apps/services do you use on this device? (for each type of device mentioned)
\end{itemize}

\noindent \textbf{Attitudes and perceived risks around data privacy} 
\begin{itemize}
\item Do you have any ideas of what types of your personal data is collected when you use these digital technologies on your devices? 
\item Considering the extend of data practices discussed just now, what are your general thoughts about data privacy in regarding digital technologies?
\item Please think about the digital tools/technologies that you use, can you think of some of the potential risks around your personal data privacy? 
\end{itemize}

\noindent \textbf{Data privacy-related information behaviors}
\begin{itemize}
\item How did you typically come across or get information about data privacy in the digital world? (follow-up questions: where, how, information sources , and perceived credibility of information sources)
\item Have you ever tried to find any information about data privacy? (If yes, follow up with where, how, and information sources; If not, ask why and hypothetical information seeking plan)
\item (If they have not mentioned privacy policies) Are you familiar with privacy policies? (follow up on perceived credibility for privacy policies)
\end{itemize}

\noindent \textbf{Needs for assistive technologies to navigate data privacy-related information} 
\begin{itemize}
\item You mentioned that you used [X]. Please imagine if there is an expert who can answer any questions around data privacy for [X]. What kinds of questions would you ask this expert about [X]? ([X] is a digital tool/technology mentioned by participants; up to 4 tools/technologies of different categories are asked here)
\item Please imagine if the expert we discussed above is a digital privacy assistant that can provide you with information around data privacy, how would you like the privacy assistant to be? [Let participants freely describe their imagined privacy assistant first; then ask follow up questions on modality, sources, and interest in using it]
\end{itemize}

\end{document}